# A Physical Model of a Smart Magneto-Composite Material and the Methodology of Research of the Elastic Properties of Nano Magnetic Composite Materials

Igor DIMITROV     Ulaş DINÇ

BAHÇEŞEHİR HIGH SCHOOL FOR SCIENCE and
TECHNOLOGY, ISTANBUL, TURKEY

# A Physical Model of a Smart Magneto-Composite Material and the Methodology of Research of the Elastic Properties of Nano Magnetic Composite Materials

**Igor DIMITROV**     **Ulaş DINC**

BAHÇEŞEHİR HIGH SCHOOL FOR SCIENCE and TECHNOLOGY,
ISTANBUL, TURKEY

# A Physical Model of a Smart Magneto-Composite Material and the Methodology of Research of the Elastic Properties of Nano Magnetic Composite Materials

## List of Illustrations

# List of Symbols

$\varepsilon$ :           energy

$\varepsilon_{ij}$ :          the energy between i-th and j-th dipole

$R_{ij}, r_{ij}$:      the distance between i-th and j-th dipoles

**m**$_i$, **m**$_j$, m:  magnetic moment

$\mu_0$:           magnetic permitivity of the free space

$\mu_B$            Bohr magneton

**B**,*b*:          magnetic field

*E,H,h*:       energy

$R_0$: the nearest neighbor distance

*a*:            the nearest neighbor distance

$\tau$:           torque

$\alpha = E/E_0$

F:              force

$\beta = a^4 E$

$\gamma = 3\beta$

*k*:            elastic constant

*keff*:         effective elastic constant

*k(a)*:         magnetic elastic constant

$k_s$:            elastic constant of a spring

**e**:           unit vector

# Abstract

In this study, we show that any system consisting of magnetic dipoles forming ordered or disordered configurations can be simplified to a form mathematically equivalent to a system consisting of two magnetic dipoles. It is shown that the energy of all kinds of magnetic dipole systems can be written as $E = -\dfrac{\beta}{a^3}$, where a is the nearest distance between the dipoles and β is a certain constant depending on the magnetic moments of dipoles and configuration. Using this fact we model any nano-magnetic composite material by a simple two-magnetic dipole system. Then we experimentaly and theoreticaly show that under certain conditions the elastic properties of the composite material can be changed using exernal magnetic field which leads to creation of smart composite materials.

# Table of Contents

# A Physical Model of a Smart Magneto-Composite Material and the Methodology of Research of the Elastic Properties of Nano Magnetic Composite Materials

## 1. Introduction

Since their synthesis in 1964 [1] suspensions of magnetic nanoparticles (diameters of the magnetic core ≈10–50 nm) in nonmagnetic carrier liquids have been called magnetic fluids (ferrofluids, ferrocolloids), and their investigation has become an independent branch of science. Particles in ferrofluids are made of Fe, Co, Ni, and their oxides. The size of the magnetic particle is smaller than the critical size of the monodomain state for the latter ferromagnetic and antiferromagnetic materials. Therefore, each particle is homogeneously magnetized. Its magnetic moment is proportional to the particle volume and the saturation magnetization of the bulk material. For nonelectrolyte carrier liquids, a steric coating of magnetic cores is used to prevent the coagulation, with an oleic acid (commonly) taken as a stabilizer [2]. Strong response to an external magnetic field, represented by ferrocolloids in combination with a liquid state, gives rise to numerous applications of magnetic fluids in engineering and natural science.

In general, if the dimension of a magnetic particle is reduced drastically, to a few nanometers, the domains in the particle merge into a single one, creating what is known as a single-domain particle and the total magnetic moment is the sum of all moments of the atomic particle. Therefore, the total magnetic moment of a single-domain nanoparticle can be about 10,000 times greater than the atomic moments of the constituent atoms. The behavior of such single-domain nanoparticles can be very complex because the magnetic moment of each particle disturbs the behavior of neighboring particles. In general, this interaction between particles serves to align the magnetic moments reducing the energy of the system. It is commonly believed that the magnetic behavior of a magnetic nano-composite material is of super paramagnetic type.

Similar to ferrofluids, there has recently been renewed interest in the magnetic properties of nano-composite materials that consist of very fine magnetic particles embedded in a metallic or dielectric nonmagnetic host [3-9] and in nano-composite materials that consist of multi-layers with very thin magnetic layers [10-11], which exhibit certain super-paramagnetic properties, too. The complexity of the magnetic properties of nano-composite magnetic materials is basically due to the dual nature of magnetic dipole-dipole interactions: if the dipoles are along a line their magnetic moment orientation is along the line (ferromagnetic ordering), however, when two dipoles are parallel then their magnetic moments are anti-parallel (anti-ferromagnetic ordering). The situation is much more complicated in a three dimensional ensemble of magnetic dipoles.

The electromagnetic properties of nano-composite materials can be very unusual compared to common materials and have been explored in many different branches of materials science engineering. At the same, it is commonly believed that the magnetic interaction is, not enough compared to the elastic forces of the host non-magnetic material in order to make significant contribution to mechanical properties of the composite material. In fact it depends on the nature of the host material, because starting with fluid host materials and going up to solid host ones, the magnetic interactions can play significant role in forming of the mechanical properties of the composite nano magnetic materials.

The aim of the present study is to show that at certain conditions, namely in the case of strong magnetic dipole interactions between the nano particles, compared with the elastic properties of the host material, magnetic interactions can contribute considerably to the elastic properties of the compound material. Though a real nano-composite magnetic material is a very complicated system from point of view of dipole-dipole interactions, any dipole system, being in its stable state, is equivalent to a single dipole system. This makes it possible to investigate the elastic behavior of a nano composite magnetic material by means of physical modeling of simple two magnetic dipole system formed by small permanent magnets and a simple string. This investigation is a preliminary theory for more complicated smart nano composite materials.

## 2. The Model of a Magnetic Dipole System

In this section, we use results obtained by different authors (for example, [12-13]) about the structure of a magnetic dipole system. The interaction energy between two point dipoles occupying lattice sites $j$, and $i$ is written as:

$$\varepsilon_{ij} = -\frac{\mu_0}{4\pi r_{ij}^3}\left[3(\mathbf{m}_j \times \mathbf{e}_{ij})(\mathbf{m}_i \times \mathbf{e}_{ij}) - (\mathbf{m}_j \times \mathbf{m}_i)\right] \tag{1}$$

Thus, the energy of the system is

$$H = -\frac{1}{2}\sum_{\substack{i,j=1\\j\neq i}}^{N}\frac{\mu_0}{4\pi r_{ij}^3}\left[3(\mathbf{m}_j \times \mathbf{e}_{ij})(\mathbf{m}_i \times \mathbf{e}_{ij}) - (\mathbf{m}_j \times \mathbf{m}_i)\right] - \mathbf{B}\sum_{i=1}^{N}\mathbf{m}_i \tag{2}$$

Here, $\mathbf{m}_i$, $\mathbf{r}_i$, $\mu_0$, $\mathbf{B}$ is the dipole moment, the position vector of the *i-th* lattice site and permeability of free space, and the external magnetic field, respectively; $r_{ij} = |\mathbf{r}_j - \mathbf{r}_i|$,

$\mathbf{e}_{ij} = \dfrac{\mathbf{r}_j - \mathbf{r}_i}{r_{ij}}$. It is convenient to present Eq,(2) in a dimensionless form: if we assume $\mathbf{m}_i = n\mu_B \mathbf{e}_i$ and $r_{ij} = r_0 R_{ij}$ with $r_0$ the nearest neighbor distance we get $H = E_0 h$ where

$$h = -\frac{1}{2}\sum_{\substack{i,j=1\\j\neq i}}^{N}\frac{1}{R_{ij}^3}\left[3(\mathbf{e}_j \times \mathbf{e}_{ij})(\mathbf{e}_i \times \mathbf{e}_{ij}) - (\mathbf{e}_j \times \mathbf{e}_i)\right] - \mathbf{b}\sum_{i=1}^{N}\mathbf{e}_i , \tag{3}$$

$$E_0 = \frac{\mu_0 \mu_B^2}{4\pi r_0^3}n^2 = \frac{\mu_0 m^2}{4\pi r_0^3}; \quad b = \frac{B\mu_B}{E_0} \tag{4}$$

$E_0$ is the energy unit, $n$ is an integer number and $\mu_B$ is the Bohr magneton. The first term in the above energy equation represents the sum of all pairs of dipole-dipole interaction energies. The second term is the energy of the aligned dipoles in the direction of the applied magnetic field. At zero temperature, the equilibrium state of the system is the state with H being at minimum. At equilibrium the torque acting on any dipole,

$$\boldsymbol{\tau}_i = \mathbf{e}_i \times \left[\sum_{\substack{j=1\\j\neq i}}^{N}\frac{1}{R_{ij}^3}\left[3\mathbf{e}_j \cdot \mathbf{e}_{ij}|\mathbf{e}_{ij} - \mathbf{e}_j\right] + \mathbf{b}\right] \tag{5}$$

must be equal to zero. It is possible to find out the minimum energy configuration of the system, i.e. to find its ground state, by using numerical methods. According to Bolcal et.al.[12], the ground state of line of diploes is equal to

$$h = -2N\sum_{j=1}^{\infty}\frac{1}{j^3} = -2.404N \tag{6}$$

thus, the energy per dipole is -2.404 (in $E_0$ units). At the same time, these authors found that the energy of a simple square lattice model or simple cubic lattice model is almost the same :-2.550 and -2.679 per dipole, respectively. For an infinitive simple square lattice model it was found that the ground state energy of the system per dipol is $\varepsilon = -2.549436$. Some of these models are shown in the figures below. According to Fig.1, the magnetic structure of the system consists of anti-ferromagnetic domains with anti-ferromagnetic domain walls.

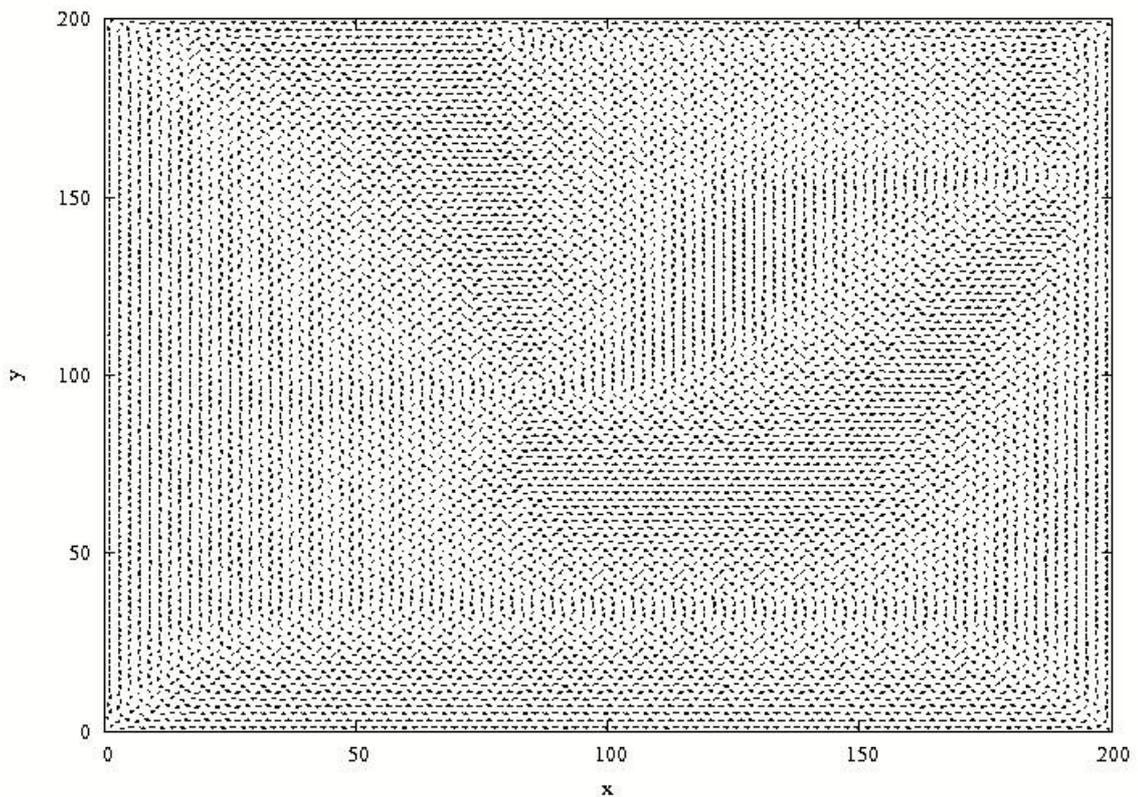

Fig.1 Projection of the dipole moments on the *x-y* plane of a 200x200 dipoles at zero temperature and in a zero magnetic field. The unit distance along the *x* and *y*-axis is the double lattice parameter of the model.

Fig.2 demonstrates the behavior of the 60x60-dipole model in the presence of an external magnetic field. In the example the magnetic field is along the *x*-direction and its magnitude b=1. Fig.3 and 4 present a 40x40 and 10x10 dipole systems, respectively, in more detailed form.

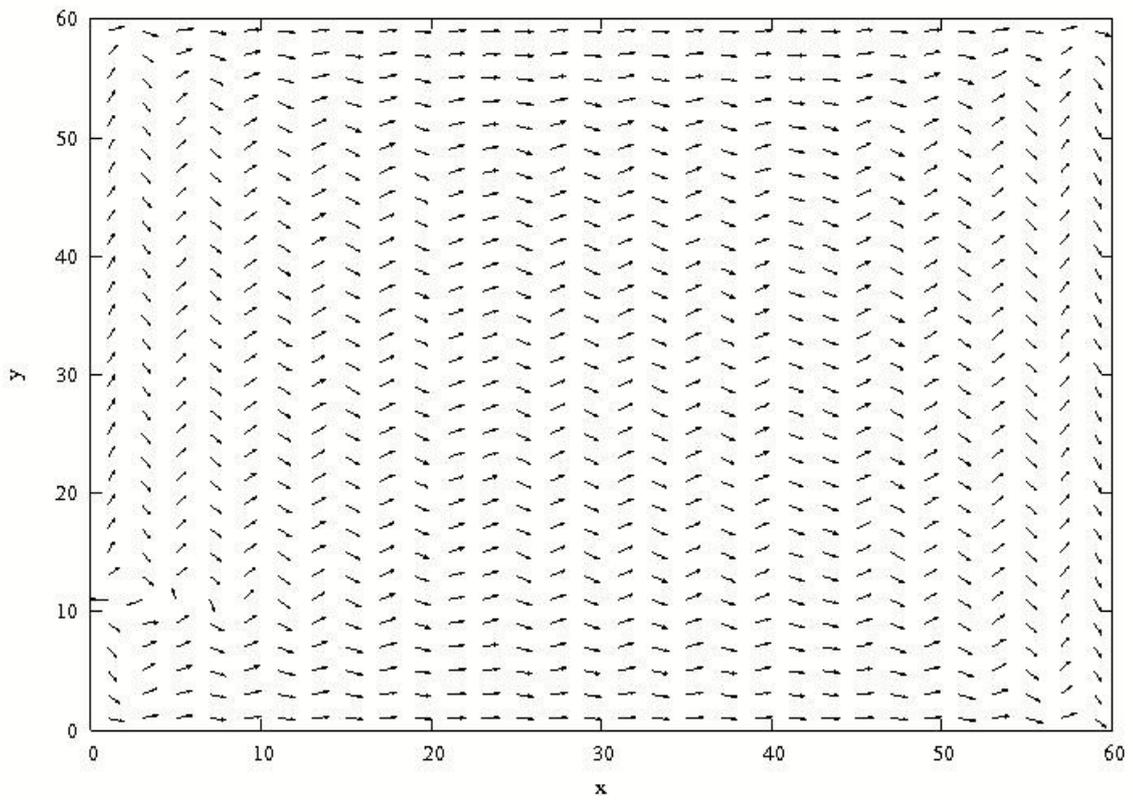

Fig.2 Projection of the dipole moments on the *x-y* plane of a 60x60 dipoles model at zero temperature and with the magnitude of magnetic field b=1, applied along the x-axis. The unit distance along the *x* and *y*-axis is the double lattice parameter of the model.

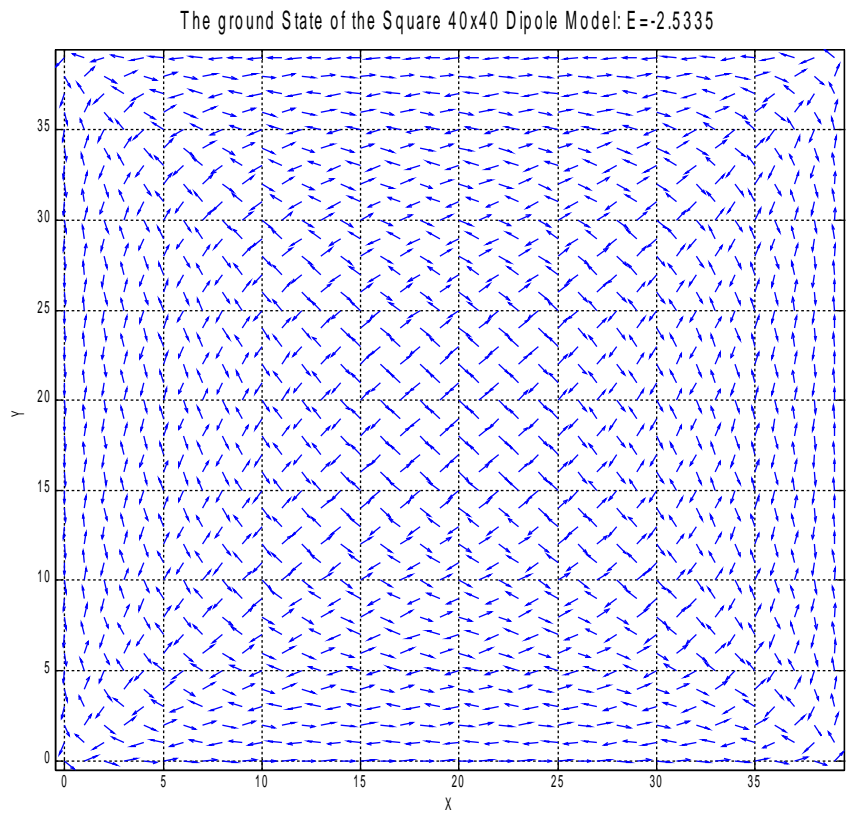

Fig. 3. The gound state of a 40x40 dipole system

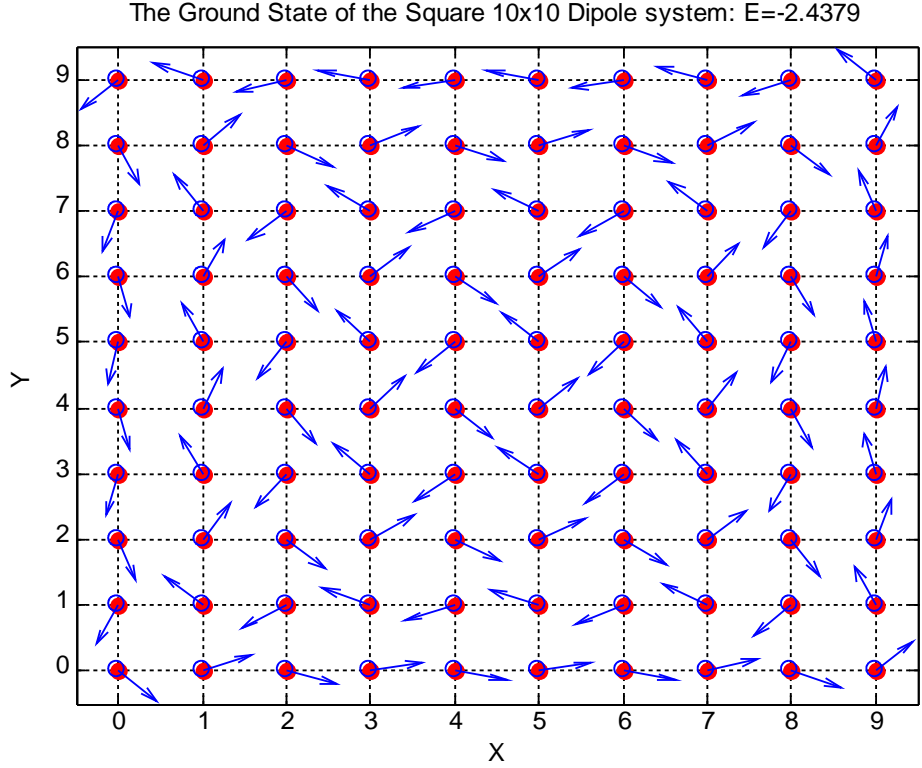

Figure 4. 10x10 magnetic dipole system at its ground state.

Here we show that even a single pair of two magnetic dipoles can model the behavior of the large dipol system. Let us now first consider the two-dipoles model the first dipole to be at the origin of the coordinate system, and the other at point (0,1), i.e.

$$\begin{cases} \mathbf{m}_1 = \mathbf{i} \\ \mathbf{m}_2 = \mathbf{e}_2 \\ \mathbf{e}_{12} = \mathbf{i} \end{cases} \quad . \tag{7}$$

According to Eq(7) the torque acting on the second dipole, is $\boldsymbol{\tau}_2 = \mathbf{e}_2 \times \mathbf{b}_2$, where $\mathbf{b}_2 = \left[ 3(\mathbf{m}_1 \cdot \mathbf{e}_{12}) \mathbf{e}_{12} - \mathbf{m}_1 \right] = \left[ 3(\mathbf{e}_1 \cdot \mathbf{i}) \mathbf{i} - \mathbf{e}_1 \right] = 2\mathbf{i}$ is the magnetic field at the position of the second dipole. Thus, the torque equals to $\boldsymbol{\tau}_2 = 2\mathbf{e}_2 \times \mathbf{i}$. Therefore, the stable state of this system is when $\boldsymbol{\tau}_2 = 2\mathbf{e}_2 \times \mathbf{i} = 0$, i.e., when $\mathbf{m}_2 = \mathbf{m}_1 = \mathbf{i}$, thus the configuration energy of the system is

$$\varepsilon = -\mathbf{m}_2 \cdot \mathbf{b}_2 = -\left[ 3(\mathbf{m}_1 \cdot \mathbf{i})(\mathbf{m}_2 \cdot \mathbf{i}) - (\mathbf{m}_1 \cdot \mathbf{m}_2) \right] = -2 \tag{8}$$

Now, let us consider a 2x2-dipole system. The configuration of the model obtained by the numerical calculations by Bolcal et.al.[12] is shown in Fig.5.

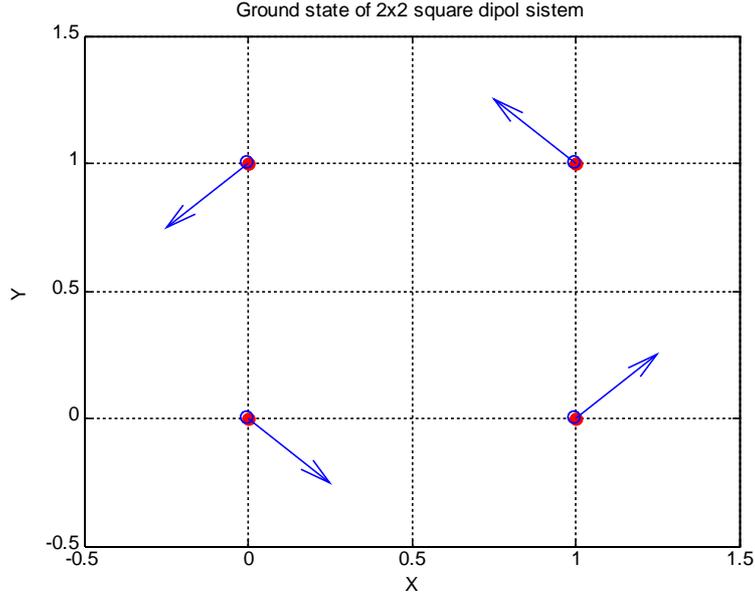

Fig. 5. The ground state configuration of the 2x2-dipol system

That this is the ground state of the model, we can show analytically in a straight way. According to Fig.5:

$$\begin{cases} \mathbf{m}_1 = \frac{1}{\sqrt{2}}(1,-1); & \mathbf{r}_1 = (0,0) \\ \mathbf{m}_2 = \frac{1}{\sqrt{2}}(1,1); & \mathbf{r}_2 = (1,0) \\ \mathbf{m}_3 = \frac{1}{\sqrt{2}}(-1,1); & \mathbf{r}_3 = (1,1) \\ \mathbf{m}_4 = \frac{1}{\sqrt{2}}(-1,-1); & \mathbf{r}_4 = (1,0) \end{cases} \quad (9)$$

Then, the magnetic field at the position of the first dipole is

$$\mathbf{b}_1 = \left( \frac{3(\mathbf{m}_2 \cdot \mathbf{r}_2)\mathbf{r}_2}{r_2^5} - \frac{\mathbf{m}_2}{r_2^3} \right) + \left( \frac{3(\mathbf{m}_3 \cdot \mathbf{r}_3)\mathbf{r}_3}{r_3^5} - \frac{\mathbf{m}_3}{r_3^3} \right) + \left( \frac{3(\mathbf{m}_4 \cdot \mathbf{r}_4)\mathbf{r}_4}{r_4^5} - \frac{\mathbf{m}_4}{r_4^3} \right) \Rightarrow$$

$$\sqrt{2}\mathbf{b}_1 = (2\mathbf{i} - \mathbf{j}) + \frac{1}{2\sqrt{2}}(\mathbf{i} - \mathbf{j}) + (\mathbf{i} - 2\mathbf{j}) = \left( \frac{1+6\sqrt{2}}{2\sqrt{2}} \right)(\mathbf{i} - \mathbf{j}) = \left( \frac{1+6\sqrt{2}}{2} \right)\mathbf{m}_1$$

Consequently,

$$\mathbf{b}_1 = \left( \frac{1+6\sqrt{2}}{2\sqrt{2}} \right)\mathbf{m}_1 \quad (10)$$

Following the symmetry:

$$\begin{cases} \mathbf{b}_1 = \left(\dfrac{1+6\sqrt{2}}{2\sqrt{2}}\right)\mathbf{m}_1 \\ \mathbf{b}_2 = \left(\dfrac{1+6\sqrt{2}}{2\sqrt{2}}\right)\mathbf{m}_2 \\ \mathbf{b}_3 = \left(\dfrac{1+6\sqrt{2}}{2\sqrt{2}}\right)\mathbf{m}_3 \\ \mathbf{b}_4 = \left(\dfrac{1+6\sqrt{2}}{2\sqrt{2}}\right)\mathbf{m}_4 \end{cases} \qquad (11)$$

Consequently, any dipole of the system is along the magnetic field that is the system is at a stable equilibrium and its energy is equal to

$$E = -\frac{1}{2}\sum_{i=1}^{4}\mathbf{m}_i \cdot \mathbf{b}_i = -2\left(\dfrac{1+6\sqrt{2}}{2\sqrt{2}}\right) = -\dfrac{1+6\sqrt{2}}{\sqrt{2}} \qquad (12)$$

Then, the energy per dipole is

$$\varepsilon = -\dfrac{1+6\sqrt{2}}{4\sqrt{2}} \approx -1.67678 \qquad (13)$$

Bolcal et. al. [12], showed that if the model size is of order of 10 units and higher, the energy per dipole of the model is quite close to that of the infinitive, i.e. it is realistic to extrapolate the properties of a finite model to that of an infinitive one. Therefore, the magnetic energy of any dipole system, per a dipole can be presented in general dimensional form as $\dfrac{E}{E_0} = -\alpha$, where $E_0 = \dfrac{\mu_0 m^2}{4\pi a^3}$, $a$ is the nearest-neighbor distance between the dipoles, and $\alpha$ is a dimensionless parameter which depends on the dimension, type, size and shape of the sample nano-magnetic composite material. It is important to underline that $\alpha$ does not differ so much for different dipole configuration and its value is nearby 2. Thus, the energy per dipole can be written as:

$$E = -\alpha\dfrac{\mu_0 m^2}{4\pi a^3} = -\dfrac{\beta}{a^3}, \qquad (14)$$

$$\text{here } \beta = \alpha \frac{\mu_0 m^2}{4\pi} = \begin{cases} 2\dfrac{\mu_0 m^2}{4\pi} & \text{- two dipoles} \\ 1.68\dfrac{\mu_0 m^2}{4\pi} & \text{- four dipoles at the corner of a square} \\ 2.404\dfrac{\mu_0 m^2}{4\pi} & \text{- infinitive line of dipoles} \\ 2.55\dfrac{\mu_0 m^2}{4\pi} & \text{- infinitive square lattice of dipoles} \\ 2.68\dfrac{\mu_0 m^2}{4\pi} & \text{- infinitive cubic lattice of dipoles} \end{cases}$$

depends on configuration of dipoles and their magnetic moments. Therefore, the elastic properties of a large magnetic dipole system can be modeled by a single pair of two dipoles, which we use in our next step of the investigation.

## 3. Methodology of Investigation of the Contribution of Magnetic Forces to Stiffness of the Composite Nano-magnetic Material

*3.1 The method of the investigation*

Here we assume that combination of the nearest distance between the magnetic nano-particles and their magnetic moment yields enough strong magnetic interaction to contribute to the mechanical properties of the compound material. As we proved in the previous section, the energy per dipole for all magnetic dipole systems can be presented in a form of Eq.(14). Therefore, we investigate the elastic properties of a simple physical model, which consists of two permanent magnets and a spring as shown in the figure below. Before going into experimental details, let us first shortly present the method of calculation of the elastic coefficient of a 'magnetic dipole string', made by two magnetic dipoles. For simplicity, we make use of only repealing magnetic interactions. According to Eq.(14) the magnitude of the force acting between two dipoles is

$$F = \frac{dE}{da} = \frac{3E}{a} = \frac{3\beta}{a^4} = \frac{\gamma}{a^4} \tag{15}$$

Thus, the magnitude of the elastic coefficient corresponding to this 'string',

$$k(a) = \frac{dF}{da} = \frac{d^2 E}{da^2} = \frac{4\gamma}{a^5} = \frac{3\alpha}{\pi}\frac{\mu_0 m^2}{a^5}, \tag{16}$$

is a function of the distance *a* between the nearest neighbor dipoles in the composite material. At the same time, the effective elastic constant of the composite material will be superposition of the elastic constants of the non-magnetic host material and that of the magnetic force. For our physical model, when the two permanent magnets repeal each other, the effective elastic coefficient will be a sum of the conventional elastic constant (that of the host material, here the string) and the magnetic one. Supposing that the magnetic interaction does not influence the elastic one, the effective elastic coefficient of the physical 'dipole-string' model well be

$$k_{eff}(a) = |k_s \pm k(a)| = \left| k_s \pm \frac{3\alpha}{\pi} \frac{\mu_0 m^2}{a^5} \right|, \qquad (17)$$

here $k_s$ is the spring elastic constant.

3.2 *The experiment*

The experiment is reduced to measuring a force: in one of the cases this is the spring force, in the other two cases, the force is magnetic or it is combination of spring and magnetic ones. The experimental set up used in all experiments is shown in Figure 6. The main part of the experimental set up is the electronic weight scale, which has a high accuracy strain-gauge sensor, providing accuracy measurement of the force within $\pm 10^{-3} N$. In order to observe the behavior of the physical model for short and long separetions of the magnets we investigated one short, 12*mm* long, and enough stiff spring, and one 17*mm* long and more soft non-magnetic spring.

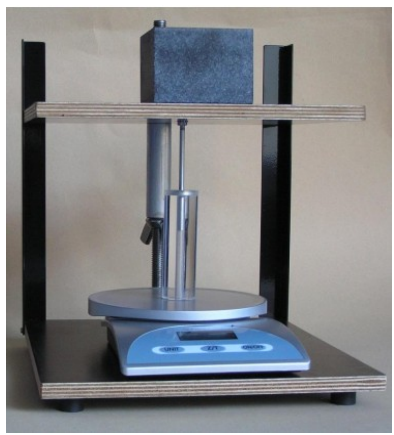

Fig.6 The experimental set up for force measuring

The schematic illustration of the experimental set up is shown in Figure 7.

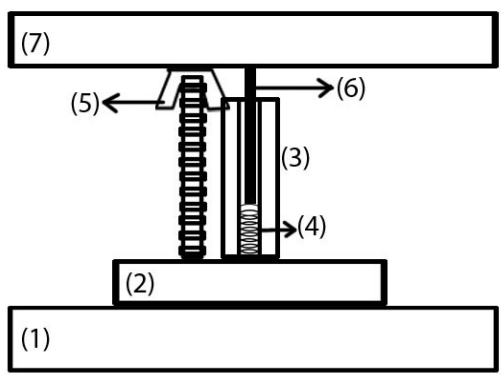

Fig.7. The sketch of the experimental set up for force measuring. The numbers indicate the following: (1)Base; (2)Scale; (3)Transparent cylinder with a small cylindrical cavity (5*mm diameter*); (4)Spring or neodymium magnets (In the figure just the spring is shown); (5)Height adjuster; (6)Small rod used for pressing spring and magnets; (7)Upper platform with adjustable height.

3.2.1 *Measuring of Elastic Constant of Springs*

First, we determine the elastic constant of the spring in the following way. When turned $180^0$, the adjuster moves 1mm along the shaft. This enables a precise control of how much spring is compressed. When the spring is compressed, the scale reads corresponding value in grams. When multiplied with gravitational acceleration we easily obtain the forces on the spring. The initial length of the spring is 12*mm*. The graph obtained from this experiment is shown in Fig.8.

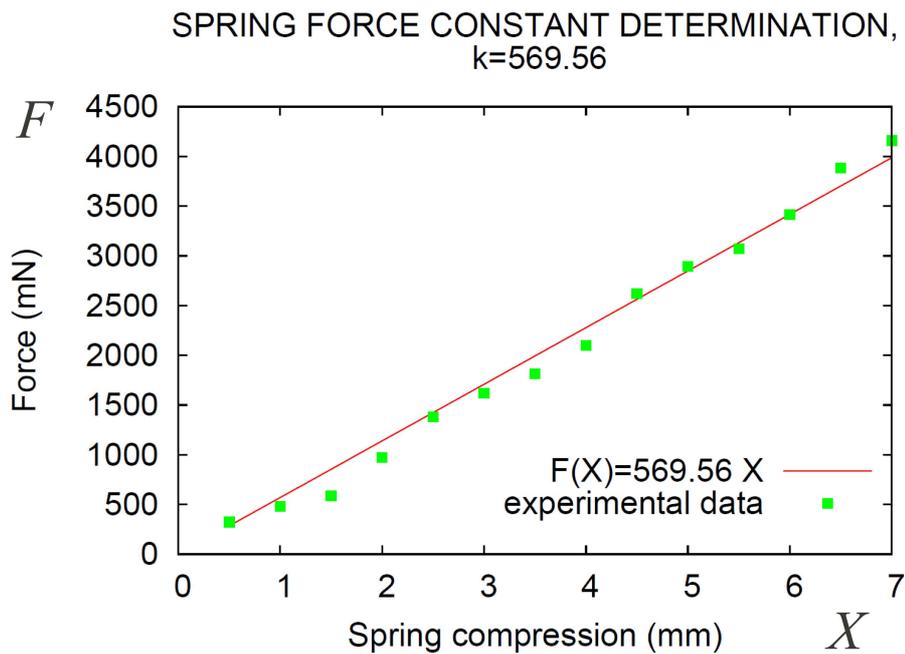

Fig.8. The spring force $F$ (in $mN$) as a function of compression, $X$, of the 12mm-long spring (in $mm$)

Calculating the average slope of of the experimental curve of the elastic force as function of the stretch of the string, we found the elastic coefficient of the spring to be equal to $k_s = 569.56 (N/m)$.

By the same way we investigated the 17mm-long soft non-magnetic spring. The result is presented in Fig.9. According to this experiment, the spring constant was calculated to be approximately $k_s = 367.4 (N/m)$

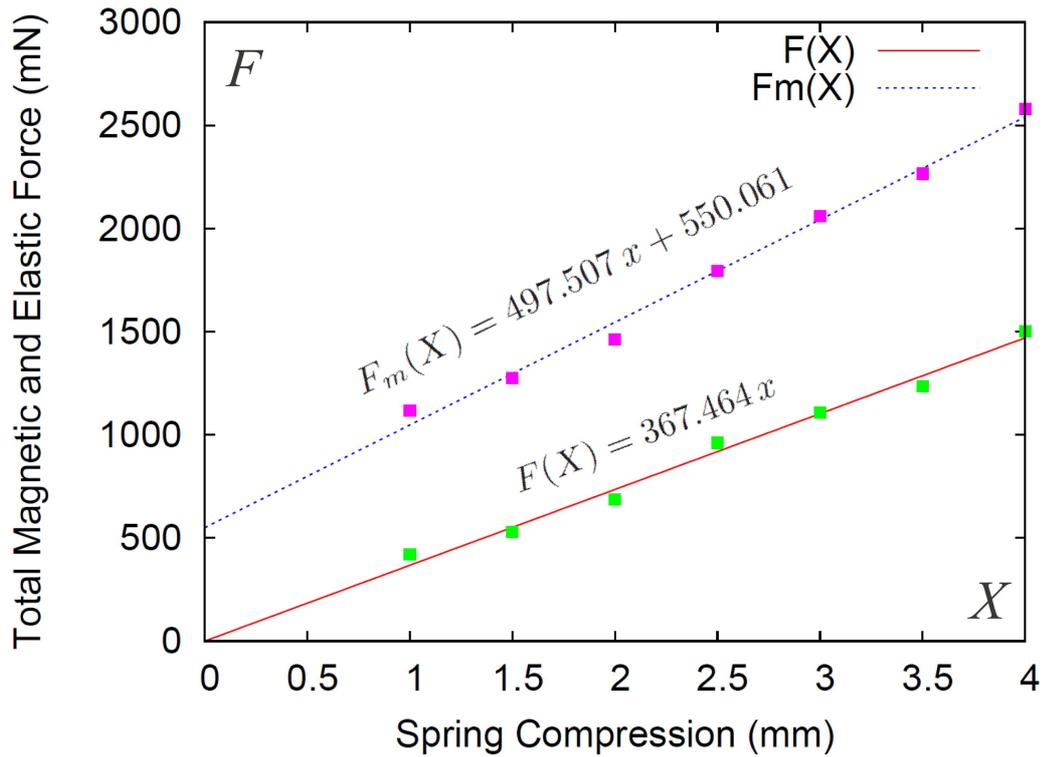

Fig.9. The spring force $F(X)$ and the total, spring plus magnetic, force $F_m(X)$ as a function of compression $X$ of the 17mm-long non-magnetic spring

3.2.2 *Measuring of Magnetic Interaction Force between the Magnets*

The next step of our experiment is to determine the law of interaction between the permanent magnets used in the investigation. The magnetic forces were measured with two neodymium magnets, one larger in size and with stronger magnetic properties. The experimental system was set up as shown in Fig.10.

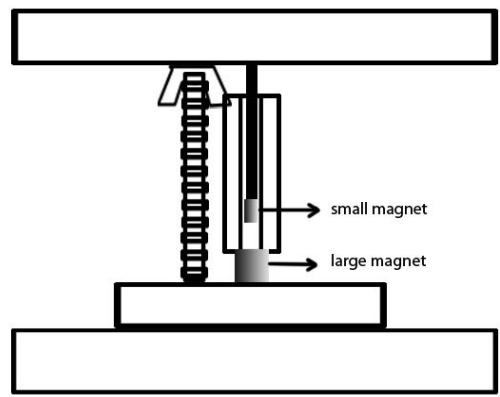

Fig.10. The sketch of the experimental set up for magnetic force measuring

The methodology in data collection remained the same as for the spring force measuring. The graph obtained from this experiment is shown in Fig.11.

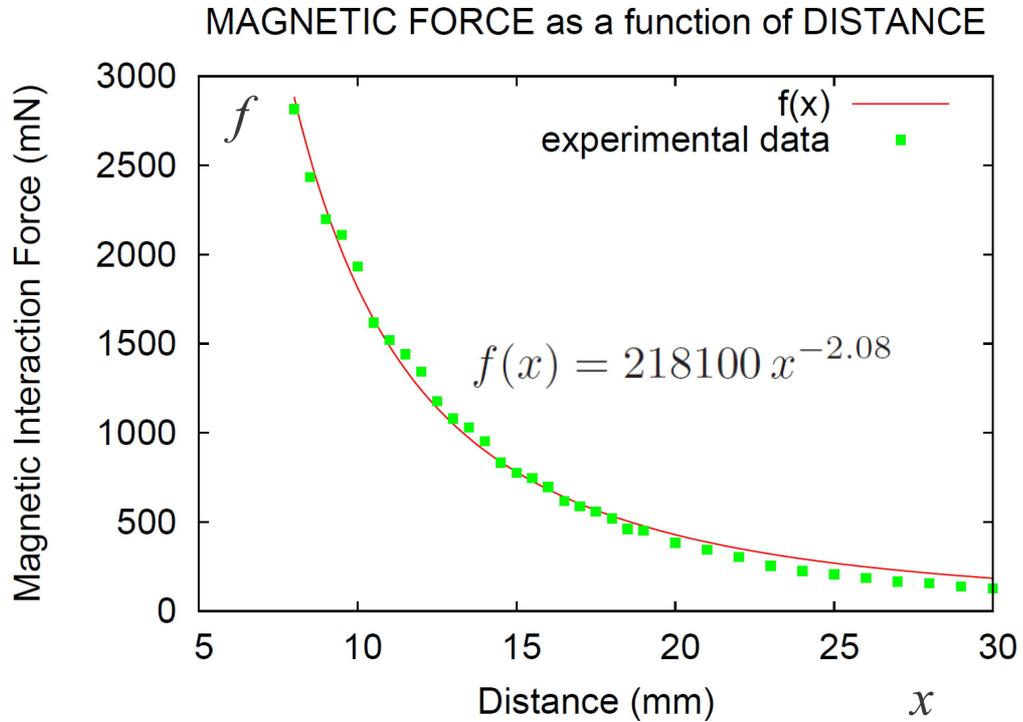

Fig.11. The magnetic force $F$ as a function of distance between magnets, $x$

At large distances, these magnets behave like magnetic dipoles, however, the interaction between them does not follow the exact law of dipole-dipole interaction, Eq.(14), because these magnets are not exactly point-like dipoles, and as a result, the law of interaction will differ from Eq.(15). If we take the data measured for both long and short distances, the law of interaction differs from Eq.(15) in the power of the exponent: for a pure dipole it is -4, while the experimental one the exponent is between -2 and -3. Therefore, we fit the experimental result according to the equation

$$\frac{F}{F_0} = \gamma \left(\frac{x}{x_0}\right)^{-n} \qquad (18)$$

where $n$ and $\beta$ are dimensionless parameters of the model to be fitted to the experimental curve, $F_0 = 9.81 \times 10^{-3} N$, $x_0 = 1mm$ are units for force and distance, respectively.

According to the experimental data the average values of these parameters are $\gamma = 218100$ and $n = 2.08$. Thus, for the magnitude of the 'magnetic' elastic constant we get

$$k(x) = \frac{dF}{dx} = k_0 n\gamma \left(\frac{x}{x_0}\right)^{-n-1} = n\gamma x^{-n-1} = n\frac{F(x)}{x}, \qquad (19)$$

here $k_0 = \frac{F_0}{x_0} = 1(N/m)$ is the unit for elastic constant, and the distance $x$ is measured in *mm*. As it should be, this constant is a function of distance. Its graph is presented in Fig.12. Accuracy of the experimentally obtained parameters $(\gamma, n,$ and $k)$ is within 8%.

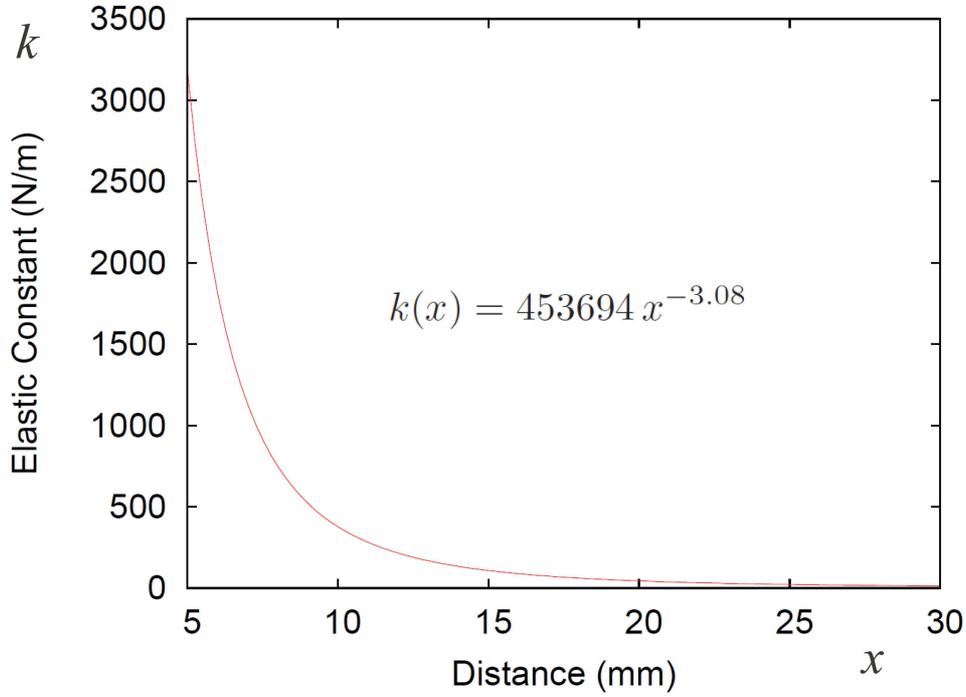

Fig.12. The magnetic force konstant $k(x) = \left|\frac{dF}{dx}\right| = n\frac{F(x)}{x}$ as a function of distance between magnets, $x$

### 3.2.3 Measuring of Magnetic and Elastic Interaction Forces in the Compound Spring-Magnets Sysytem

In this section, we investigate the elastic properties of the compound string-magnets system, which models the real nano-composite magnetic materials. In the model, the most important problem is whether the principle of superposition of elastic and magnetic forces works. The principle of adding the magnetic and elastic forces in the experiment is obeyed much better for non-magnetic springs then for magnetic ones. This is so, because the magnetic springs, due to their magnetization, contribute to some extend to the interactions in the magnet-spring system. Therefore, we check the superposing of the elastic and magnetic forces on the non-magnetic spring. For this, we put the non-magnetic spring between the magnets and measure the force acting on the weight scale. The experiment done confirmed the superposition of the elastic and magnetic forces. For example, when the separation between the magnets was 14*mm* (i.e. the compression of the spring was 3*mm*), the measured force was 1590(*mN*). For the same distance, according to Fig.9, the elastic force is 990(*mN*), while, the calculated magnetic force is approximately 530(mN), i.e. quite good consistence is observed for the principle of superposition of elastic and magnetic forces. Similar result is observed for other compressions of the nonmagnetic spring. Unfortunately, with the available nonmagnetic spring, we could not go beyond the minimum (12*mm*) separation between magnets because this distance corresponded to the maximum compressing of the spring used in the experiment, as well as there was an upper limitation for the separation, which was equal to the free length of the spring, equal to 17 *mm*. In this range for separation of magnets the slope of the magnetic curve cannot be calculated with a good accuracy (the force varies slowly for these separations), while force itself is measured enough accurate. The experimental curve obtained for the force as a function of the compression of the spring-magnets system is shown in Fig.9. In this region of compression of the system the contribution of the magnetic forces to the effective elastic constant is considerable, about 35%.

**4. Results and Discussions**

We have proposed a realistic physical model for investigation of the elastic properties of a composite ferromagnetic material. The physical model consists of two magnetic dipoles and a string. The physical model represents adequately the stiffness properties of a real nano magnetic composite material because a system consisting of large number of dipoles mathematically is equivalent with a system of two magnetic dipoles only. In our experiment, instead of a real magnetic dipole, which magnetic field is presented in Fig13.a, we have used a short cylindrical magnet, which magnetic field is shown in Fig.13.b. The reason is that more massive is the magnet the grater is the force of interaction and the higher accuracy of the experiment.

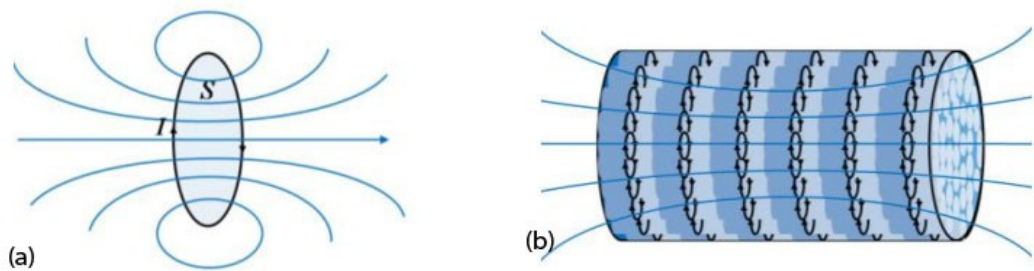

Fig. 13. The magnetic field of a magnetic dipole and a short cylindrical permanent magnet

From Fig.13, it becomes evident that the shapes of these magnetic systems are mathematically different, therefore the force of acting between the magnets used in our experiment does not follow the form of the magnetic force of dipole, Eq.(15). However, when the distance between the magnets becomes some multiples of the size of the magnets, the law of magnetic interactions tends to that of Eq.(15), i.e. the power of the exponents start tending towards 4.

The contribution of the magnetic interactions to the stiffness of the magneto-composite materials expressed trough the effective stiffness coefficient, $k_{eff}(a) = k_s + k(a) = k_s + \frac{4\gamma}{a^5}$, depends basically on the nearest distance between the dipoles and the coefficient $\gamma = 3\alpha \frac{\mu_0 m^2}{4\pi}$. This coefficient accounts for many factors, such as dimension, size, and shape of the composite material, as well as the magnitude of the magnetic moments of diploes constituting the compound material. If the magnetic forces were forces of attraction then the effective elastic coefficient were $k_{eff}(a) = |k_s - k(a)| = \left|k_s - \frac{4\beta}{a^5}\right|$. Thus, varying the above-mentioned factors it is possible to design a composite material of desired stiffness. Applying of external magnetic field will change parameter $\alpha$ responsible for the magnetic configuration of the dipoles, which will change mechanical properties of the material. Indeed, the stiffness of the host material is very important, for example if the stiffness of the host material is much grater of the contribution of the magnetic force, as it is true for very hard solid materials, then controlling the mechanical properties of a composite material does not work. However, if the host material is soft comparing to magnetic interaction, as it is valid for ferrofluids, then controlling the mechanical properties via the magnetic field would become a powerful tool for creation of smart composite materials. In our physical model the combination of the used magnets and the soft spring leads to the class of soft host materials in the compound, the increase of the elastic constant was about 35%.

## 5. Conclusion

The elastic properties of nano-magnetic composite materials can be modeled by a pair of two magnetic dipoles separated by a string representing the elastic properties of the host non-magnetic material. Depending on the elastic properties of the host material, the magnetic dipole-dipole interactions can significantly change the elastic properties of the compound materials especially when the host material is not stiff enough. For these types of materials, the elastic properties of the compound can be modified by an applied external magnetic field. The magnetic interactions between the magnets, used in our physical model, changed the elastic constant of the spring used in the experiment about 35%. This contribution can be varied by using different magnets or springs.

When designing a smart magneto-composite material we should first calculate the stiffness constant of the magnetic dipole forces following the expression for the stiffness of the magnetic dipole forces, Eq.(16):

$$k(a) = \frac{3\alpha}{\pi} \frac{\mu_0 m^2}{a^5} \approx 2 \frac{\mu_0 m^2}{a^5}$$

If the combination between the magnetic moment of nano particles ($m$) and the nearest distance ($a$) between them gives stiffness comparable of that of the host material ($k_s$), then it is possible to design a smart material. Because, by applying magnetic field we can change the configuration of the dipole system, thus to change coefficient $\alpha$ in Eq.(16) and getting desirable effective stiffness coefficient $k_{eff}(a) = \left| k_s \pm \frac{3\alpha}{\pi} \frac{\mu_0 m^2}{a^5} \right|$ for the compound material. This coefficient can be done even zero, like for liquids.